\documentclass[%
  prl,
 amsmath,amssymb,
reprint,%
]{revtex4-1}
\usepackage{graphicx}% Include figure files
\usepackage{dcolumn}% Align table columns on decimal point
\usepackage{bm}% bold math
\usepackage{color}
\usepackage{float}

\usepackage[utf8]{inputenc}
\usepackage[T1]{fontenc}
\usepackage{mathptmx}
\usepackage{subfig}
\usepackage[inline]{enumitem}
\usepackage{caption}
\newcommand{\bogus}[1]{{}}
\begin{document}

%\preprint{AIP/123-QED}

%\preprint{AIP/123-QED}

\title{Cooling flow regime of a plasma thermal quench}

\author{Yanzeng Zhang}%
\affiliation{Theoretical Division, Los Alamos National Laboratory, Los Alamos, New Mexico 87545, USA}
\author{Jun Li}%
\affiliation{Theoretical Division, Los Alamos National Laboratory, Los Alamos, New Mexico 87545, USA}
\affiliation{School of Nuclear Science and Technology, University of Science and Technology of China, Hefei, Anhui, China}
\author{Xian-Zhu Tang}%
\affiliation{Theoretical Division, Los Alamos National Laboratory, Los Alamos, New Mexico 87545, USA}


\begin{abstract}
 A large class of Laboratory, Space, and Astrophysical plasmas is
  nearly collisionless. When a localized energy or particle sink, for
  example, in the form of a radiative cooling spot or a black hole, is
  introduced into such a plasma, it can trigger a plasma thermal
  collapse, also known as a thermal quench in tokamak fusion.  Here we
  show that the electron thermal conduction in such a nearly
  collisionless plasma follows the convective energy transport scaling
  in itself or in its spatial gradient, due to the constraint of
  ambipolar transport. As the result, a robust cooling flow aggregates
  mass toward the cooling spot and the thermal collapse of the
  surrounding plasma takes the form of four propagating fronts that
  originate from the radiative cooling spot, along the magnetic field
  line in a magnetized plasma. The slowest one, which is responsible
  for deep cooling, is a shock front.

\end{abstract}

\maketitle

\maketitle A signature property of a large class of magnetized and
unmagnetized plasmas in the Laboratory, Space, and Astrophysical
systems is the extremely low collisionality that can be due to high
plasma temperature $T_e$ or low plasma density $n_e,$ or a combination
of the two~\cite{NAP-2010}. For example, a fusion-grade plasma in a
tokamak reactor has $T_e\sim 10-20$ kilo-electron-volts (KeV) and $n_e
\sim 10^{19-20}$ per cubic meter, which result in a mean-free-path
$\lambda_{mfp} \sim 10^4$~meters (m), while the toroidal length of the
confinement chamber is merely
20-30~m~\cite{ITER-transport-nf-2007,federici-etal-fed-2018,Fernandez-etal-nf-2022}. In
the earth's radiation belt, the electron $\lambda_{mfp}$ can be as
long as $10^{11}$~m with electron energy from tens of KeV to MeV and
an $n_e\sim
10^4$~m$^{-3}$~\cite{Parks-ency-atom-science-2015,denton-etal-JGRSP-2010,denton-cayton-AG-2011,borovsky-etal-JGRSP-2016}.
At the even grander scale of clusters of galaxies, the intracluster
hot gas has $n_e\sim 10^{2} - 10^{4}$~m$^{-3}$ and $T_e\sim 2\times
10^{7}-10^{8}$~K~\cite{Fabian-ARAA-1994,fabian-springer-2002,Peterson-Fabian-PR-2006,Hitomi-collaboration-nature-2016,Zhuravleva-etal-nature-2014},
so $\lambda_{mfp}$ is in the order of tens of kilo-parsec to
mega-parsec.
 
A whole class of problems arises if a localized cooling spot is
introduced into such a nearly collisionless plasma. This could be
structure formation in a galaxy cluster where a radiative cooling spot
is driven by increased particle density~\cite{Fabian-ARAA-1994} or an
event horizon of a black hole that provides an absorbing boundary for
plasmas~\cite{krolik-book-1999}.  A satellite
traversing the earth's radiation belt can be a sink for plasma energy
and
particles~\cite{hastings-garrett-book-1996,Nwankwo-etal-BookChapter-2020}.
In a tokamak reactor, solid pellets that are injected into the fusion
plasma for fueling and disruption
mitigation~\cite{federici2003key,baylor2009pellet,combs2010alternative},
provide localized cooling due to a combination of energy spent on
phase transition and ionization of the solid materials, and the
radiative cooling that is especially strong when high-Z impurities are
embedded in the frozen pellet.  Even in the absence of pellet
injection, large-scale magnetohydrodynamic instabilities can turn
nested flux surfaces into globally stochastic field lines that connect
fusion core plasma directly onto the divertor/first
wall~\cite{bondeson1991mhd,riccardo2002disruption,nardon2016progress,sweeney2018relationship},
causing a thermal collapse via fast parallel transport along the field
lines within a short period of time that can range from micro-seconds
to milliseconds~\cite{riccardo2005timescale}.  An outstanding physics
question is how a thermal collapse of the surrounding plasma, commonly
known as a thermal quench in tokamak fusion, would come about in such
a diverse range of applications.

The most obvious route for the thermal collapse is via electron 
thermal conduction along the magnetic field line that intercepts the 
cooling spot, for which the Braginskii formula~\cite{braginskii} would 
produce an enormous heat 
flux~\cite{Fabian-ARAA-1994,Peterson-Fabian-PR-2006} if there is a 
sizeable temperature difference $\Delta T = T_0 - T_w \sim T_0$ 
between the cooling spot ($T_w$) and the surrounding plasma ($T_0$), 
\begin{align} 
  q_{e\parallel} = n_e \chi_e \mathbf{\hat{b}} \cdot \nabla T_e \sim 
  n_e v_{th,e} \frac{\lambda_{mfp}}{L_T} \Delta 
  T \sim n_e v_{th,e} \frac{\lambda_{mfp}}{L_T} T_0. \label{eq:qe-bragisnkii} 
\end{align} 
Here $v_{th,e}=\sqrt{T_0/m_e}$ is the electron thermal speed and $L_T$
the distance or field line length over which the temperature drop
$\Delta T$ is established. For a nearly collisionless plasma, the
temperature collapse necessarily starts with Knudsen number $K_n
\equiv \lambda_{mfp}/L_T \gg 1,$ a regime in which the free-streaming
limit~\cite{atzeni_book_2004} of
\begin{align} 
q_{e\parallel} \approx \alpha n_e v_{th,e} T_0, \label{eq:qe-flux-limiter} 
\end{align}
is supposed to apply in lieu of Braginskii, with $\alpha \approx 0.1$~\cite{Bell-pof-1985}.

The pressure-gradient-driven plasma flow $V_{i\parallel}$ along the 
magnetic field line is limited by the ion sound speed $c_s,$ so the 
convective electron energy flux is bounded by $n_e c_s T_0.$ The 
scaling of $q_{e\parallel}$ with $v_{th,e}$ in 
Eq.~(\ref{eq:qe-flux-limiter}) suggests that the electron energy flux 
would be dominated by conduction as normally $v_{th,e} \gg c_s$ in a 
plasma of comparable electron and ion temperatures. In such a 
conduction-dominated situation, the much colder but denser cooling 
spot would be rapidly heated up by the electron thermal conduction 
from the surrounding hot plasma, and as the result, it can become 
over-pressured and the original cooling spot, say an ablated pellet in 
a tokamak, tends to expand into the surrounding plasma, yielding an 
outflow. 
 
In the aforementioned problem of clusters of galaxies, one has instead
observed robust cooling flows into the radiative cooling spot that
aggregate mass onto the cooling spot~\cite{Fabian-ARAA-1994}, although
more recent observations reveal a more modest mass-accreting cooling
flow that indicates the role of various additional heating mechanisms
to balance the
cooling~\cite{Peterson-Fabian-PR-2006,Hitomi-collaboration-nature-2016,Zhuravleva-etal-nature-2014}.
This is inconsistent with the conduction-dominated scenario mentioned
above~\cite{Binney-Cowie-apj-1981,Fabian-etal-AAR-1991,Fabian-ARAA-1994,Peterson-Fabian-PR-2006}.
Extensive efforts have been made to find ways to inhibit the electron
thermal conduction in the nearly collisionless plasma, for example, by
tangled magnetic
fields~\cite{tribble-mnras-1989,chandran-cowley-prl-1998} or plasma
instabilities~\cite{Jafelice-aj-1992,Balbus-Reynolds-apj-2008,Roberg-Clark-etal-prl-2018},
in order to reach the convection-dominated scenario, which would
naturally yield the {\em cooling flow regime} of a plasma thermal
quench.
 
In this Letter, we show that in a nearly collisionless plasma, even
along the magnetic field lines, like in the case of pellet injection
into a tokamak, ambipolar transport will naturally constrain the
electron parallel thermal conduction in such a way that the plasma
thermal collapse comes with a cooling flow toward the radiative
cooling spot.  The necessary constraint is on the spatial gradient of
electron parallel conduction flux, which can be seen from the energy
equation for the electrons along the magnetic field,
\begin{align}
  n_e\left(\frac{\partial}{\partial t} T_{e\parallel} +
  V_{e\parallel}\frac{\partial}{\partial x} T_{e\parallel}\right) +
  2n_e T_{e\parallel}\frac{\partial}{\partial x}V_{e\parallel} +
  \frac{\partial}{\partial x} q_{en} = 0.
\end{align}
Here $x$ is the distance along the magnetic field line, $n_e,
T_{e\parallel}, V_{e\parallel}$ are the density, parallel temperature,
and parallel flow of the electrons, and $q_{en} \equiv \int m_e
\left(\mathbf{v}_\parallel - V_{e\parallel}\right)^3d^3\mathbf{v}$ is
a component of the parallel heat flux.  Let the cooling flow span a
length $L_T,$ one can see the convective energy transport terms follow
the scaling of $n_e T_{e\parallel} V_{e\parallel}/L_T.$ Ambipolar transport
constrains $V_{e\parallel}\approx V_{i\parallel} \propto m_i^{-1/2},$ so
the free-streaming scaling of $q_{en}$ in
Eq.~(\ref{eq:qe-flux-limiter}) would predict $\partial q_{en}/\partial
x \sim \alpha n_e v_{th,e} T_{e\parallel}/L_T \propto m_e^{-1/2},$
which would overwhelm the convective energy transport ($\propto
m_i^{-1/2}$) to force a $T_{e\parallel}$ collapse and remove the
pressure gradient drive that sustains the cooling flow.  The condition
for accessing the cooling flow regime of plasma thermal quench is thus
\begin{align}
\frac{\partial q_{en}}{\partial x} \sim n_e T_{e\parallel}
V_{i\parallel}/L_T. \label{eq:qen-x-gradient}
\end{align}
We report in this Letter that this is indeed realized by ambipolar
constraint in a nearly collisionless plasma. In the case that the
cooling spot is a perfect particle and energy sink (e.g., a black hole),
which can be modeled by an absorbing boundary, $q_{en}$ itself obtains
the convective energy transport scaling, $q_{en} \sim n_e
V_{i\parallel}T_{e\parallel}.$ With a radiative cooling mass, which
can be modeled as a thermobath, the boundary of which recycles all
particles across the boundary but clamps the temperature to a low
value $T_w\ll T_0,$ the cold electrons thus produced can restore the
free-stream scaling for $q_{en} \sim \alpha n_e v_{th,e}
T_{e\parallel}$ but its spatial gradient over the cooling flow region retains
the convective energy transport scaling of
Eq.~(\ref{eq:qen-x-gradient}).  As the result, a robust cooling flow
appears to aggregate mass towards the cooling spot.  

\begin{figure}[hbt]
\centering
\includegraphics[width=0.45\textwidth]{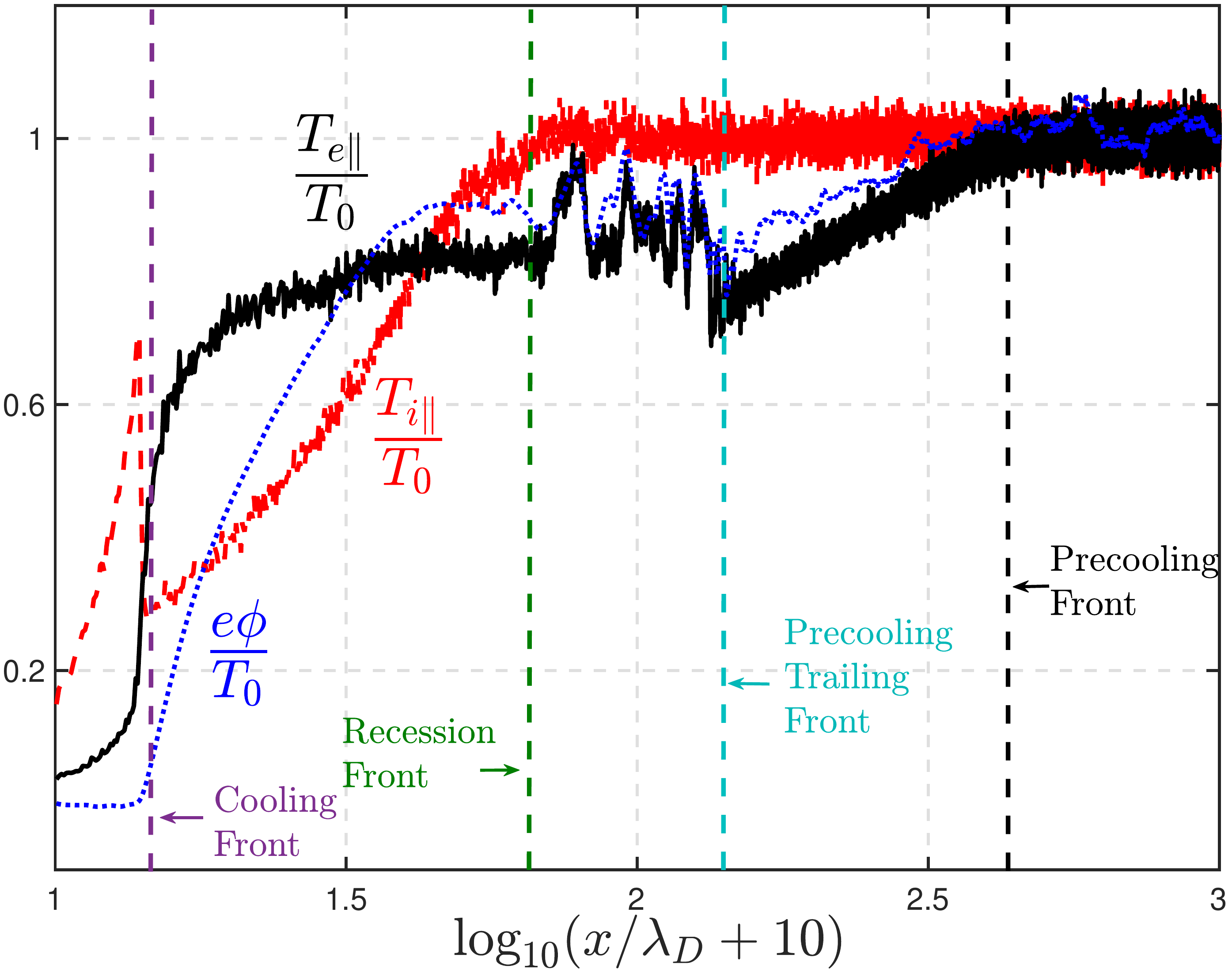}
\caption{Normalized parallel electron and ion temperature, and
  electrostatic potential at $\omega_{pe}t=176$ for
  $T_w=0.01T_0$ from first-principle VPIC~\cite{VPIC} simulation. Different fronts are labeled, where a uniform plasma
  with constant temperature $T_0$ and density $n_0$ initially fills the whole
  domain and a reduced ion mass of $m_i=100m_e$ is
  utilized. $\lambda_D$ is the Debye length.}
\label{fig:diagram}
\end{figure}

Most interestingly, in such a cooling flow regime, the plasma thermal
collapse comes in the form of propagating fronts that originate from
the cooling spot with characteristic speeds. There are totally four
(three) propagating fronts for the thermobath (absorbing) boundary:
two of them propagate at speeds that scale with $v_{th,e},$ so are
named {\em electron fronts}, while the other two are {\em ion fronts}
that propagate at speeds that scale with the local ion sound speed
$c_s$ (the last ion front disappears for the absorbing
boundary). Fig.~\ref{fig:diagram} illustrates the structure of the
four fronts that propagate into a hot plasma for the thermal collapse
with a thermobath boundary. It is important to note that cooling of a
nearly collisionless plasma produces strong temperature anisotropy, so
one must examine the collapse of $T_{\parallel}$ and $T_{\perp}$
separately.
 
Cooling of $T_{e\parallel}$ in a nearly collisionless plasma is
primarily through free-streaming loss of suprathermal electrons
satisfying $v_\parallel < - \sqrt{2e(\Delta\Phi)_{\rm max}/m_e}$ into
the radiative cooling spot. Here $(\Delta\Phi)_{\rm max}$ is the
maximum reflective potential in the plasma with $\Delta\Phi =
\Phi_\infty - \Phi(x)$ and the constant $\Phi_\infty$ the far upstream
plasma potential. The precooling zone bounded by the precooling front
(PF) and the precooling trailing front (PTF) has
$T_{i\parallel}$ unchanged and $V_{i\parallel}\approx 0,$ but a
lowered $T_{e\parallel},$ which is due to the depletion of fast
electrons satisfying $v_\parallel >v_c = \sqrt{2 e
  \left[\left(\Delta\Phi\right)_{max} - \Delta\Phi\right]/m_e}$, yielding a truncated
Maxwellian of the form
\begin{align} 
f_e(v_\parallel, v_\perp) =& \frac{n_m\left(\Phi(x)\right)}{\sqrt{2\pi} 
  v_{th,e}^3} e^{-\left(v_\parallel^2+v_\perp^2\right)/2v_{th,e}^2} 
\Theta\left(1-\frac{v_\parallel}{v_c}\right) \label{eq:truncated-fe} \\\nonumber
&+ \frac{n_b}{2\pi v_\perp}\delta(v_\perp)\delta(v_\parallel - v_c),
\end{align} 
with $\Theta(1-v_\parallel/v_c)$ the Heaviside step function that
vanishes for $v_\parallel > v_c,$ and $\delta(x)$ the Dirac delta
function. The ambipolar electric field can draw some low-energy
electrons to compensate for the loss of high-energy electrons and thus
maintain quasi-neutrality.  This in-falling cold electron population is
modeled in Eq.~(\ref{eq:truncated-fe}) as a cold beam that due to
ambipolar electric field acceleration has the speed $v_c$. Between the PF and PTF, the electron beam can be ignored, so
\begin{equation} 
  T_{e\parallel}(v_c) = \frac{\int m_e \tilde{\mathbf{v}}_\parallel^2 
    f_ed\mathbf{v}}{ \int f_e d\mathbf{v}} \approx 
  T_0\left[1-\frac{v_{c}}{\sqrt{2\pi}v_{th,e}}e^{-{v_c^2}/{2v_{th,e}^2}}\right],\label{eq:T-vc-cutoff}
\end{equation} 
 where $\tilde{\mathbf{v}}_\parallel\equiv\mathbf{v}_\parallel - \mathbf{V}_{e\parallel}$. 

Eq.~(\ref{eq:T-vc-cutoff}) predicts a detectable decrease in
$T_{e\parallel}(v_c)$ from $T_0$ for $v_c \approx 2.4 v_{th,e},$
suggesting an electron PF propagating at
\begin{equation} 
U_{PF}=2.4 v_{th,e}.
\end{equation} 
This corresponds to fast electrons with 
$v_\parallel>U_{PF}$ traveling from the left boundary into the plasma, leaving 
behind a distribution at the PF with a void in 
$v_\parallel>U_{PF}.$ The PTF comes about due to 
the reflecting potential $(\Delta\Phi)_{\rm RF}=(\Delta\Phi)_{\rm max} 
- (\Phi_\infty-\Phi_{\rm RF})$ with $\Phi_{\rm RF}$ the ambipolar 
potential at the ion recession front (RF), which sets a lower cutoff speed 
$v_c$ at 
\begin{equation} 
U_{PTF} \equiv \sqrt{2e\left(\Delta\Phi\right)_{\rm 
    RF}/m_e}. 
\end{equation} 
The deeper void now gives rise to a further reduced 
$T_{e\parallel},$ 
\begin{equation} 
  T_{e\parallel}(U_{PTF}) \approx T_0 \left[1- 
    \sqrt{{e\left(\Delta\Phi\right)_{\rm RF}}/{\pi T_0}} e^{- 
      {e\left(\Delta\Phi\right)_{\rm RF}}/{T_0}} \right]. 
\end{equation} 
The PTF rides these electrons that are reflected 
by the reflecting potential, and propagates at $U_{PTF} (< U_{PF}).$ 
Since the ambipolar reflecting potential must satisfy 
$e\left(\Delta\Phi\right)_{\rm RF} \sim T_0$ in a nearly collisionless 
plasma, $U_{PTF} \sim v_{th,e}$ and $T_{e\parallel}(U_{PTF})$ is only 
mildly cooler than $T_0.$ Furthermore, $T_{e\parallel}$ and $\Phi$ {\em vary little} 
between the RF and PTF, since the cutoff velocity remains the same at $U_{PTF}.$ 
 
The ion flow remains vanishingly small 
ahead of the RF, so the electron cooling between the 
RF and PF is the result of 
electron conduction, which for the model $f_e$ in Eq.~(\ref{eq:truncated-fe}) with 
$v_c > v_{th,e}$ takes the form 
\begin{align} 
  q_{en}\equiv \int m_e \mathbf{\tilde{v}}_\parallel^3 f_e d\mathbf{v} 
  \approx & - \frac{n_m v_{th,e} T_0}{\sqrt{2\pi}} 
  \left(\frac{v_c^2}{v_{th,e}^2}-1\right) e^{-v_c^2/2v_{th,e}^2}\label{eq-qen}\\\nonumber
  &+n_bT_0\frac{v_c^3}{v_{th,e}^2}. 
\end{align} 
Between PTF and PF, $n_b\approx 0$ and $U_{PTF} \le v_c \le U_{PF}$ so $q_{en}$ does
scale as the free-streaming limit of Eq.~(\ref{eq:qe-flux-limiter}),
but with $\alpha$ modulating in space as a function of $v_c.$
In fact, for $v_c > \sqrt{2}v_{th,e},$ one finds 
\begin{equation} 
\frac{d q_{en}}{dx} \approx  n_m v_c \frac{\partial T_{e\parallel}}{\partial x}, 
\end{equation} 
so the solution of the energy equation, $\partial 
T_{e\parallel}/\partial t = -v_c \partial T_{e\parallel}/\partial x,$ 
reveals that $v_c$ is the recession speed of $T_{e\parallel},$ 
re-affirming the particle picture noted earlier that the momentum space void in $f_e$ 
propagates upstream with a speed of $v_c.$
This large $q_{en}$ drives fast propagating electron fronts
(PF and PTF) but produces modest amount of $T_{e\parallel}$ cooling
for the large cutoff speed $v_c=U_{PTF}.$

Much more aggressive cooling would need to occur as the plasma
approaches the radiative cooling spot that is clamped at $T_w\ll T_0.$
These are facilitated by the ion fronts that provide the reflecting
potential $\left(\Delta\Phi\right)_{\rm RF}.$ The RF is where $n_i\approx n_e$ starts to drop, and behind which plasma
pressure gradient drives a cooling flow toward the radiative
cooling spot. The main reflection potential, which is tied to the
electron pressure gradient, is also behind the ion RF. An
ion recession layer bounded by the RF and the cooling front (CF)
is similar to the rarefaction wave formed in the cold plasma
interaction with a solid
surface~\cite{allen1970note,cipolla1981temporal},
where the plasma parameters recede steadily with the local sound
speed. What is different for the thermal quench of a nearly
collisionless plasma is the large plasma temperature and pressure
gradient and the nature of the heat flux. The electron flow associated
with $f_e$ of Eq.~(\ref{eq:truncated-fe}) within the recession layer
is
\begin{align}
n_eV_{e\parallel}(v_c) = - \frac{n_m v_{th,e}}{\sqrt{2\pi}}
e^{-v_c^2/2v_{th,e}^2} + n_b v_c, \label{eq-Vepara}
\end{align}
with $ n_e(v_c) = \left[1+{\rm
    Erf}\left(v_c/\sqrt{2}v_{th,e}\right)\right] n_m/2 + n_b.  $ For
an absorbing boundary ($n_b=0$), a cutoff speed around $v_{th,e}, v_c
\approx v_{th,e} \sqrt{2\ln (v_{th,e}/V_{i\parallel})},$ is sufficient to produce
a $V_{e\parallel}$ that matches onto the increasing ion flow,
$V_{e\parallel}\approx V_{i\parallel},$ for ambipolar transport
through the recession layer.  The in-falling cold electron beam reduces
 $v_c$ and hence produces a lower reflecting potential across
the recession layer as elucidated in Eq.~(\ref{eq-Vepara}).

\begin{figure}[h]
\centering
\includegraphics[width=0.45\textwidth]{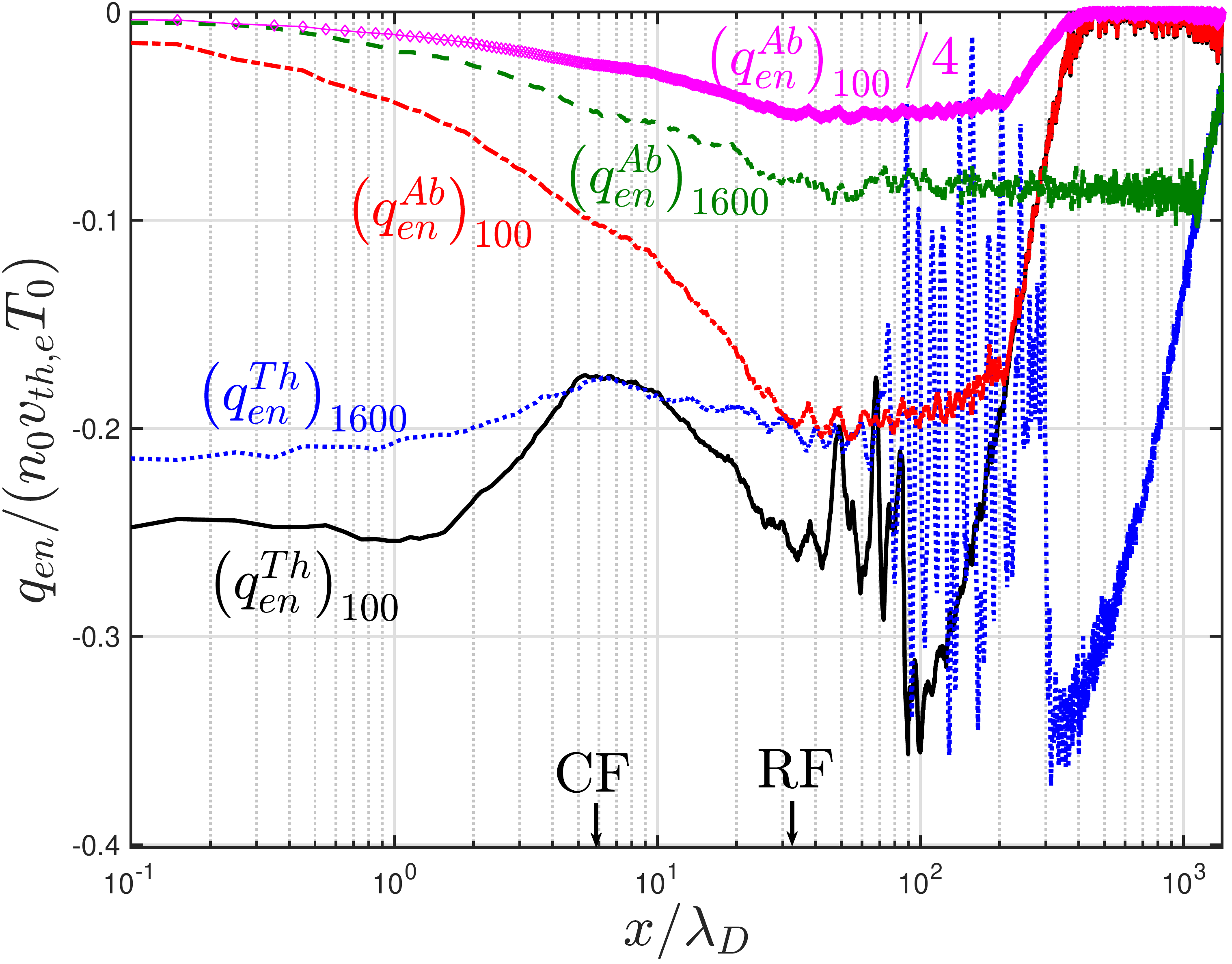}
\caption{Electron heat flux $q_{en}$ for different
  ion-electron mass ratios and boundary conditions at
  $\omega_{pi}t=13.6$. The superscripts $Ab$ and $Th$ denote the
  absorbing and thermobath boundary, respectively, while the
  subscripts $100$ and $1600$ represent $m_i/m_e=100$ and $1600$. For
  the absorbing boundary, $q_{en}$ itself behind the recession front
  (RF) follows the convective scaling, which can be seen by comparing
  $\left(q_{en}^{Ab}\right)_{1600}$ and
  $\left(q_{en}^{Ab}\right)_{100}/4$ (notice that their small
  difference, as seen from Eq.~\ref{eq:qen-coldbeam} for $n_b=0$,
  comes from the dependence of $v_c \approx v_{th,e} \sqrt{2\ln
    (v_{th,e}/V_{i\parallel})}$ on $m_i$). For the thermobath
  boundary, $q_{en}$ recovers the free-streaming formula, but its
  spatial gradient within the cooling flow region, which is between
  the cooling front (CF) and the RF, follows the convective scaling,
  which is illustrated by the same slope of the curve as that for the absorbing
  boundary when $m_i/m_e$ is fixed. }
\label{fig:heat-flux}
\end{figure}

The physics of $q_{en}$ in the {\em recession layer} can be
elucidated by rewriting Eq.~(\ref{eq-qen}) as
\begin{align}
q_{en} = \left(\frac{v_c^2}{v_{th,e}^2}+2\right)n_eV_{e\parallel}T_0-2n_bv_cT_0
  - 3n_e T_{e\parallel} V_{e\parallel} + 2 n_em_eV_{e\parallel}^3. \label{eq:qen-coldbeam} 
\end{align}
For an absorbing wall, $n_b=0,$ and one finds $q_{en}$
itself has a convective energy transport scaling: $q_{en} \sim n_e
V_{e\parallel} T_0.$ The condition for the cooling flow regime,
Eq.~(\ref{eq:qen-x-gradient}), is obviously satisfied.  In the case of
a radiative cooling spot that produces copious amount of cold
electrons, the leading order of $q_{en} \sim - 2n_b v_c T_0 \propto
n_b v_{th,e} T_0$ follows the free-streaming limit of
Eq.~(\ref{eq:qe-flux-limiter}).  Remarkably the plasma thermal quench
still produces a cooling flow, in which case
Eq.~(\ref{eq:qen-x-gradient}) is satisfied due to the collisionless
cold beam in the ambipolar electric field follows flux conservation
$n_bv_c = constant,$ so $\partial (- 2n_b v_c T_0)/\partial x=0$ and
the remaining terms in $q_{en}$ have convective energy transport
scaling.  The VPIC~\cite{VPIC} kinetic simulations shown in
Fig.~\ref{fig:heat-flux} confirms that convective scaling of
Eq.~(\ref{eq:qen-x-gradient}) holds in the recession
layer. In other words, electron cooling in a
nearly collisionless plasma is modified by ambipolarity in such a way
that large $T_{e\parallel}$ gradient can be supported in the recession
layer to drive a cooling flow.

The propagation speed of the RF can be understood by
examining the ion dynamics in the recession layer~\cite{guo-tang-pop-2012a,Li-etal-prl-2022}
\begin{align} 
 \frac{\partial }{\partial t}n_{i} + \frac{\partial}{\partial x} 
 \left(n_{i} V_{i\parallel}\right) & = 0, \label{eq-ion-density} \\
    m_{i} n_{i}\left(\frac{\partial }{\partial t} V_{i\parallel}+ 
   V_{i\parallel}\frac{\partial}{\partial x} V_{i\parallel} \right) + 
   \frac{\partial }{\partial x} (p_{i\parallel}+p_{e\parallel}) & = 
   0,\label{eq-ion-momentum} \\
 n_i \left( \frac{\partial}{\partial t} T_{i\parallel} +V_{i\parallel} 
 \frac{\partial}{\partial x} T_{i\parallel}\right)+ 
 2n_iT_{i\parallel}\frac{\partial}{\partial x}V_{i\parallel} 
 +\frac{\partial}{\partial x} q_{in} & = 0,\label{eq-ion-temperature} 
\end{align} 
where we
invoked the electron force balance $e n_{e} E_\parallel\approx -
\partial p_{e\parallel}/\partial x$ and quasi-neutrality $n_e=Zn_i$
with $Z$ the ion charge, and $p_{i,e\parallel} =
n_{i,e}T_{i,e\parallel}$. Introducing a parameterization of $q_{in}\approx \sigma_i
n_i V_{i\parallel} T_{i\parallel},$ which is known from
Ref.~\onlinecite{tang2016kinetic}, and $\partial q_{en}/\partial x \approx
\sigma_e\partial\left( n_e V_{e\parallel}
T_{e\parallel}\right)/\partial x$ from Eq.~(\ref{eq:qen-x-gradient}), we obtain an universal length scale for $p_{e,i\parallel}$
\begin{equation} 
\frac{d\textup{ln}p_{e\parallel}}{dx}\approx \mu 
\frac{d\textup{ln}p_{i\parallel}}{dx},\label{eq-dpe-dx-Z} 
\end{equation} 
where $\mu=(3+\sigma_e)/(3+\sigma_i) \times 
[-U+(1+\sigma_i)V_{i\parallel}]/[-U+(1+\sigma_e)V_{i\parallel}]$. It
is interesting to note that $U>0$ and $V_{i\parallel}<0$ have opposite
sign in the recession layer where a cooling flow resides. As a result, Eqs.~(\ref{eq-ion-density}-\ref{eq-ion-temperature}) have self-similar
solutions with similarity variable
$\xi=x-Ut$ with $U$ being the local recession speed. We find
\begin{align} 
U=\left[\sigma_i^2V_{i\parallel}^2/4+(1+\sigma_i/3)c_s^2\right]^{1/2}+(1+\sigma_i/2)V_{i\parallel},\label{eq-U} 
\end{align} 
where $c_s=\sqrt{3(\mu ZT_{e\parallel}+T_{i\parallel})/m_i}$ is the
local sound speed of a nearly collisionless plasma with anisotropic
temperatures. At the ion
recession front, $V_{i\parallel}\approx 0,$ so the speed of the ion
recession front is
\begin{equation} 
U_{RF}=\sqrt{1+\sigma_i/3} c_s. \label{eq:U-RF} 
\end{equation} 
For $Z=1$, $\sigma_i=1$, $\mu=1$ and $T_{e\parallel}\approx
T_{i\parallel}=T_0$ at the recession front, we have $U_{RF}\approx
2.8v_{th,i}$ with $v_{th,i}=\sqrt{T_0/m_i}$ the ion thermal speed,
which agrees well with the simulation result. It is worth noting that
the self-similar solution of Eq.~(\ref{eq-U}) also recovers a known
constraint~\cite{tang2016critical} on the plasma exit flow at an absorbing boundary where a
non-neutral sheath would form next to it as shown in the Supplement
material.

\begin{figure}[h]
\centering
\includegraphics[width=0.45\textwidth]{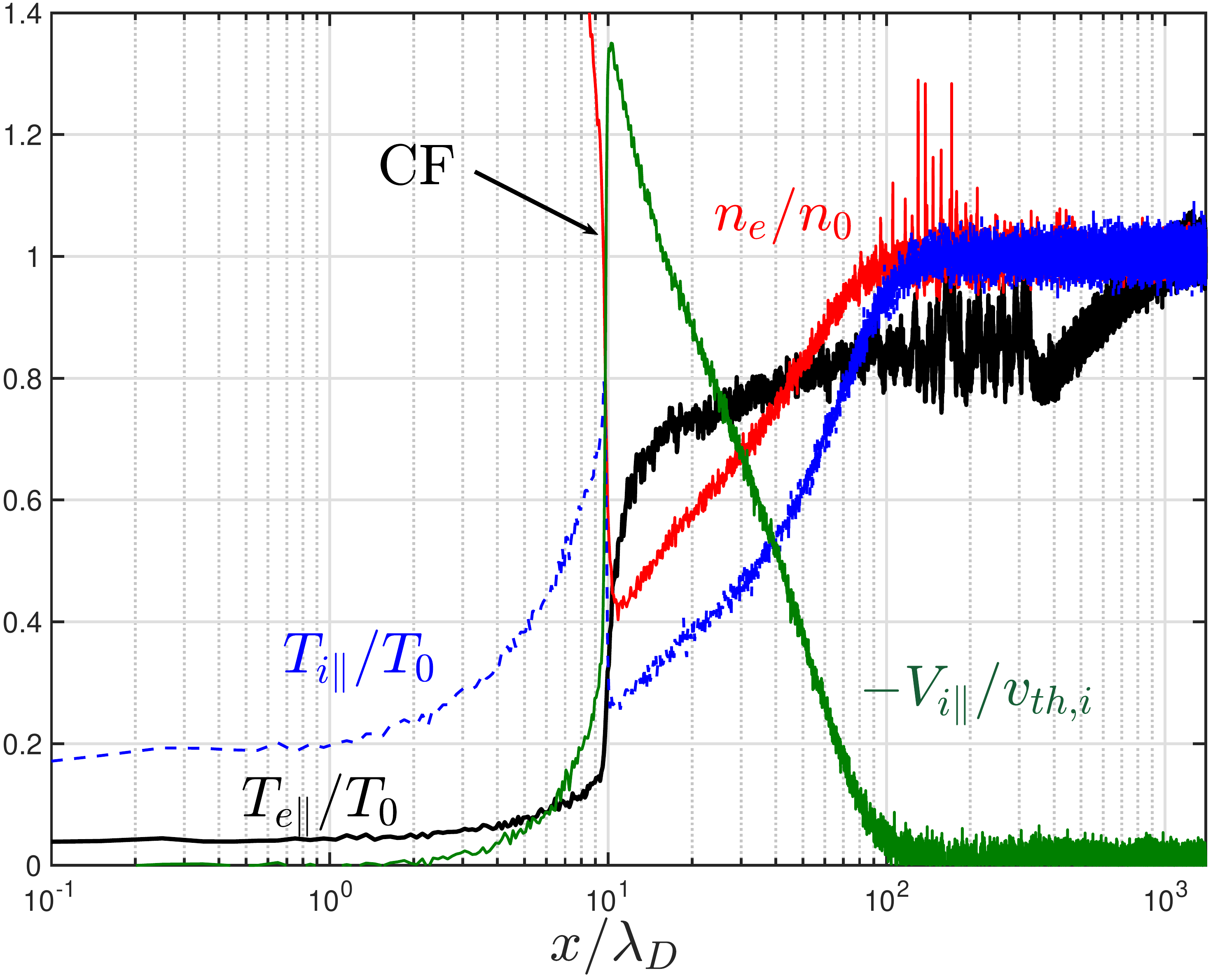}
\caption{Plasma profiles corresponding to
  Fig.~\ref{fig:diagram}. The jumps near the cooling front (CF) are
  illustrated, where the huge plasma density near the radiative
  cooling boundary $n_e>n_0$ is cut out off the figure.}
\label{fig:jump_CF}
\end{figure}

In the absence of an absorbing boundary, the mass aggregated by the
cooling flow will pile up, and the resulting back-pressure can now
drive a second ion front (cooling front, CF). Behind the CF, $T_{e\parallel}$ equilibrates with $T_w$ as shown in
Fig.~\ref{fig:jump_CF}. Such a deep cooling of $T_{e\parallel}$ is
through thermal conduction as indicated in
Fig.~\ref{fig:heat-flux}. When the cooling flow runs into this nearly
static plasma, the ion flow energy, which is substantial in the
cooling flow, is converted into ion thermal energy via a plasma shock
as shown in Fig.~\ref{fig:jump_CF}. Matching the conserved quantities
across the shock while ignoring the heat flux, we find that the speed
of the shock, which propagates upstream into the plasma, is simply the
upstream sound speed at the shock front. The CF is the shock front, so
its speed is
\begin{equation} 
U_{CF} = c_s(x=x_{CF}). 
\end{equation} 
Since the plasma temperature at the CF is considerably lower than that
at the RF, we have $U_{CF} < U_{RF}.$ Generally, the colder $T_w,$
the smaller $U_{CF}.$ The presence of the CF and the cooling zone
behind it, is of fundamental importance to $T_{i\perp}$ and
$T_{e\perp}$ cooling as the cold particles provide dilutional
cooling. It is also the source of cold electrons that are accelerated
by the ambipolar electric field into the recession layer and beyond,
cooling down $T_{e\perp}$ further upstream.
 
In conclusion, the thermal collapse of a nearly collisionless plasma
due to its interaction with a localized particle or energy sink, is
associated with a cooling flow toward the cooling spot. This applies
to unmagnetized plasmas, for example, in astrophysical systems, and
magnetized plasmas, for example, in earth's magnetosphere or a tokamak
fusion plasma. It is the fundamental constraint of ambipolar
transport, along the field line in a magnetized plasma, that limits
the spatial gradient of electron (parallel) heat flux to the much
weaker convective ($V_{e\parallel}$) scaling as opposed to the
free-streaming ($v_{th,e}$) scaling.  Such weaker scaling is essential
to sustain a temperature and hence pressure gradient for driving the
cooling flow toward the cooling spot over the ion recession layer.
The cooling flow eventually terminates against the cooling spot via a
plasma shock that converts the ion flow energy into ion thermal
energy. This shock or cooling front propagates away from the cooling
spot at upstream ion sound speed, and it has the most profound role in
the deep cooling of the surrounding hot plasmas, especially the ions.
Unlike the ions, the electrons can be cooled ahead of the recession
front due an electron heat flux that follows the free-streaming limit
($q_{en} \propto n_e v_{th,e} T_e$). Interestingly this large heat
flux does not imply significant cooling of $T_{e\parallel}$ in a
nearly collisionless plasma ahead of the recession front, but induces
a very limited amount of $T_{e\parallel}$ drop over a very large
volume, because the precooling and precooling trailing fronts have
propagation speeds that scale with electron thermal speed.
 
We thank the U.S. Department of Energy Office of Fusion Energy
Sciences and Office of Advanced Scientific Computing Research
for support under the Tokamak Disruption Simulation (TDS) Scientific
Discovery through Advanced Computing (SciDAC) project, and the Base
Theory Program, both at Los Alamos National Laboratory (LANL) under
contract No. 89233218CNA000001. Y.Z. is supported under a
Director’s Postdoctoral Fellowship at LANL. This research used
resources of the National Energy Research Scientific Computing Center
(NERSC), a U.S. Department of Energy Office of Science User Facility
operated under Contract No. DE-AC02-05CH11231 and the Los Alamos
National Laboratory Institutional Computing Program, which is
supported by the U.S. Department of Energy National Nuclear Security
Administration under Contract No. 89233218CNA000001.

\bibliography{reference}% common bib file

\end{document}